\documentclass[aps,twocolumn,prb,reprint,amsmath,amssymb,showpacs,superscriptaddress]{revtex4-1}
\usepackage{multirow}
\usepackage{graphicx,epsfig}
\usepackage{wasysym}
\usepackage{color,soul} 
\usepackage[colorlinks=true,citecolor=blue,linkcolor=red,linktocpage=true,pagebackref=false]{hyperref}
\usepackage{cleveref}
\usepackage[linktocpage=true]{hyperref}
\usepackage{booktabs}

\newcommand{\ba}{\begin{eqnarray}}
\newcommand{\ea}{  \end{eqnarray}}

\def \beq {\begin{equation}}
\def \edq {\end{equation}}
\def \bes {\begin{subequations}}
\def \eds {\end{subequations}}
\def \beqn {\begin{equation*}}
\def \edqn {\end{equation*}}

\begin{document}
\title{Anderson-Mott transition in a disordered Hubbard chain with correlated hopping.}
\author{Francesca Battista}
\affiliation{Departamento de F\'{\i}sica, FCEyN, Universidad de Buenos Aires and IFIBA, Pabell\'on I, Ciudad Universitaria, 1428 CABA Argentina}
\author{Alberto Camjayi}
\affiliation{Departamento de F\'{\i}sica, FCEyN, Universidad de Buenos Aires and IFIBA, Pabell\'on I, Ciudad Universitaria, 1428 CABA Argentina}
\author{Liliana Arrachea}
\affiliation{Departamento de F\'{\i}sica, FCEyN, Universidad de Buenos Aires and IFIBA, Pabell\'on I, Ciudad Universitaria, 1428 CABA Argentina}
\affiliation{International Center for Advanced Studies, ECyT-UNSAM,
Campus Miguelete, 25 de Mayo y Francia, 1650 Buenos Aires, Argentina}

\begin{abstract}
 We study the ground state phase diagram of the Anderson-Hubbard model with correlated hopping at half filling in one-dimension. The Hamiltonian has a local Coulomb repulsion $U$ and 
 a disorder potential with local energies randomly distributed in the interval $(-W,+W)$ with equal probability, acting on the singly occupied sites.
The hopping process which modifies the number of doubly occupied sites is forbidden.
 The hopping between nearest-neighbor singly occupied and empty sites or between singly occupied and doubly occupied sites have the same amplitude $t$. 
We identify three different phases as functions of the disorder amplitude
$W$ and Coulomb interaction strength $U>0$. When $U<4t$ the system shows a metallic phase (i) only when no disorder is present $W=0$ or an Anderson-localized phase (ii) when disorder is introduced $W\neq 0$. When $U>4t$ the Anderson-localized phase survives as long as disorder effects dominates on the interaction effects, otherwise a Mott insulator phase (iii)  arises.  The phases (i) and (ii) are characterized by a finite density of doublons and a vanishing charge gap between the ground state and the excited states. 
The phase (iii) is characterized by vanishing density of doublons and
a finite  gap for the charge excitations.
\end{abstract}

\pacs{72.15.Rn 05.30.Rt 05.30.Fk}
\maketitle
\section{Introduction}

The idea of inducing electron localization with many-body interactions in fermionic systems was introduced by Mott in the framework of a crystalline array with strong Coulomb interactions.\cite{mott,mott1,mott2} 
As in the case of usual band-insulators, a gap opens in the spectrum of the many-particle system. In this case the origin of this  feature is  the effect of the interactions rather than  the nature of the atomic structure.
The Hubbard model constitutes the tight-binding version of this picture. The corresponding Hamiltonian  has a hopping term between nearest neighbors in a lattice  combined with a  local Coulomb interaction. \cite{hub1,hub2,hub3}
The competition between the kinetic energy and the strong correlations represented by these two terms is precisely the origin of the metal-insulator transition proposed by Mott. For this reason, the Hubbard 
model was thought to be the natural candidate to realize the Mott transition \cite{mott1}. In bipartite lattices, however, the nesting in the Fermi surface  and the relevant umklapp  processes at half-filling lead to magnetic
instabilities that
mask the metallic phase. The ground state is insulating for any value of the Coulomb interaction and, unlike the original Mott picture, it is dominated by strong antiferromagnetic correlations.
 \cite{imada} When the antiferromagnetism is frustrated, several results support the scenario 
 of the metal-insulator Mott transition in this model. This is the case of  lattices with frustrating geometries, \cite{krish} infinite-range hopping or lattices with an infinite coordination number. \cite{dinf} 

The Anderson transition defines a completely different paradigm of   metal-insulator transition.  The model  was introduced at the end of the 50's 
\cite{and} to describe the localization of  non-interacting single particles due to the introduction of disorder in a crystal potential landscape. The Hamiltonian has
 a hopping term between nearest-neighbor sites in a lattice with randomly distributed local energies.
 Unlike the Hubbard model, the insulating phase is gapless. In this phase
 the electrons get trapped and localized in real space due to the potential landscape \cite{thou, Kram,  lee, eve}, while
 the metallic phase is stabilized for small strength of disorder in 3D lattices.
 In lower dimensionality, the ground state is always insulating. \cite{2d, 1d}

The combination of Coulomb interaction and local potential disorder defines the disordered Hubbard model and the outcome of such an interplay is highly non-trivial. \cite{lee,belkir,cas,lange,krz} 
The naive expectation that the combination of two ingredients leading to  insulating phases also results in an insulating phase does not apply to this model. In fact, quantum Monte Carlo results 
 in 2D\cite{sri, dent, chak} and results obtained with dynamical mean field theory  in lattices with infinite connectivity  \cite{jan-vol,ulm,krz} suggest the possibility of a metal-insulator transition in the phase diagram of the Coulomb interaction $U$ vs the strength of disorder $W$ at half-filling.
 The concept of  many-body localization has been recently coined to characterize the effect of disorder in the presence of many-body interactions  and is receiving lot of attention for some years now \cite{basko, oga, pal, berkel, luitz, bera}. Theoretical studies include perturbative calculations \cite{fleish, alts, jac} and numerical simulations. \cite{oga, pal, ser, can, cueva} These ideas are also motivating   
 experimental studies, not only for solid state systems but also in other correlated systems like cold atoms and optical lattices.\cite{kondo, schre}
 In fact, the advances in quantum optics enabled the experimental realization of optical potentials imitating a crystal lattice (optical lattices)\cite{deut}, with the advantage of having tunable parameters. \cite{tie, chin, lye, falla} For example, experimental studies of a Hubbard model in three dimension have been done  using ultracold neutral atoms trapped in an optical lattice.\cite{jak,white} The interplay of interaction and disorder is crucial when studying cold atoms in these lattices too. \cite{billy, roati, white, sanchez}

\begin{figure}[h!]
\begin{center} 
\includegraphics[width=0.5\textwidth]{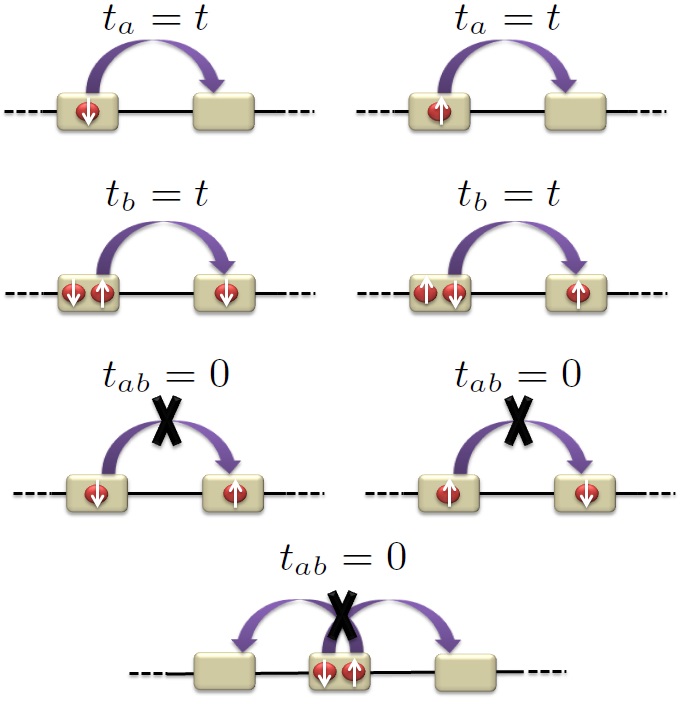}
\caption{ (Color online) Schematic representation of the possible hopping events considered in this paper. The hopping takes place between a singly occupied site and an empty site with amplitude $t_a=t$ or between a single occupied site and a doubly occupied site with amplitude $t_b=t$. The hopping between two singly occupied sites with different spins component to generate a doubly occupied site and an empty site or vice-versa is forbidden $t_{ab}=0$.
}\label{hopscheme}
\end{center}
\end{figure}

The Hubbard model with correlated hopping supports a limit with the basic ingredients for a Mott transition.
The corresponding Hamiltonian has a local Coulomb interaction $U$ identical to the one of the usual Hubbard model, but the kinetic term is generalized to have different amplitudes ($t_{a}$, $t_{ab}$, $t_b$) 
depending on the occupation of the two neighboring sites $\langle ij \rangle$ that intervene in the hopping process with a total number of $1, 2, 3$ particles, respectively. The corresponding kinetic term reads
 \begin{eqnarray}\label{cor-hop}
& &  H(t_a,t_{ab}, t_b)= \sum_{\langle ij \rangle, \sigma}^{L}c_{i, \sigma}^{\dagger}c_{j,\sigma}\{ t_a ~(1-n_{i,-\sigma})(1-n_{j,-\sigma}) \nonumber \\
 & &~~~~~~~~+t_{ab} ~\left[n_{i,-\sigma}(1-n_{j,-\sigma}) + (1-n_{i,-\sigma}) n_{j,-\sigma}  \right] \nonumber \\
 & & ~~~~~~~~+ t_b ~ n_{i,-\sigma}n_{j,-\sigma}\} +h.c.
\end{eqnarray}
The operators $c_{i, \sigma}^{\dagger} ~ (c_{i,\sigma})$ create (annihilate) a $S=1/2$- fermion with spin component $\sigma=\uparrow, \downarrow$ at the chain site $i$, and 
$n_{i,\sigma}=c_{i, \sigma}^{\dagger}c_{i,\sigma}$ is the corresponding number operator. 
The hopping process that permutes a singly occupied site and an empty site has amplitude $t_a$. The one  permuting a single occupied site and a doubly occupied site  has amplitude $t_b$. The hopping
between two singly occupied sites with opposite spins to generate an empty site and a doubly occupied site or vice-versa has amplitude $t_{ab}$ (see the sketch of Fig. \ref{hopscheme}).
These terms emerge naturally in the 
derivation of an effective low-energy model starting from the three-band Hubbard  Hamiltonian that describes the Cu-O planes of  the superconducting cuprates. \cite{sim,sup1,sup2} In some limit, it is equivalent to the so called extended Hubbard model with bond-charge interaction which has been investigated in the context of low-dimensional organic superconductors \cite{bond-ch1,bond-ch2,bond-ch3} as well as in the context of  non-conventional superconducting mechanisms, like hole-superconductivity \cite{hir1,hir2} and $\eta$-pairing superconductivity, \cite{sch,mont3} mesoscopic transport, \cite{mes}  and quantum information. \cite{mont1,mont2} More recently, this model was also investigated in  optical lattices and cold atoms \cite{optlat0,optlat1,optlat2,optlat3,optlat4,optlat5} and was also found to provide the effective theory of the Hubbard Hamiltonian in driven lattices. \cite{driv1,driv2}  

The correlated hopping Hubbard Hamiltonian can be exactly solved in 1D in two limits. One corresponds to  the usual Hubbard model where $t_a=t_b=t_{ab}$, which can be solved by Bethe ansatz. \cite{lieb-wu}
The other solvable case is  $|t_a|=|t_b|=t$ and $t_{ab}=0$.\cite{aa1,aa2} 
At half-filling, when the Coulomb interaction $U$ overcomes a critical value $U_c$, the ground state corresponds to an  insulator, with an energy gap increasing linearly with $U$. It is interesting to notice that for $t_{ab}=0$ the antiferromagnetic correlations are completed inhibited. Hence, the insulating phase has the characteristic of an ideal Mott insulator, in the sense that it does not have any magnetic order.
Below the critical value $U_c$, the system is gapless and has the characteristics of a normal metal. 

The aim of this paper is to investigate the phase diagram of the correlated Hubbard model in the exactly solvable limit of Ref. \onlinecite{aa1} at half-filling with  an additional term in the Hamiltonian that represents a disordered potential for the singly occupied sites. 
The model is introduced in Section \ref{model}. In Section \ref{method} we present the methodology to investigate the ground state of this Hamiltonian. Results are presented in Section \ref{results} and Section \ref{conc} is devoted to summary and conclusions.

\section{Model}\label{model}
We study a disordered Hubbard model with correlated hopping. The corresponding Hamiltonian is
 \begin{equation}
 H=H_K+H_U
 \end{equation}
 where
 \begin{eqnarray}
 H_K&=&H(t,0,t)+
 \sum_{i, \sigma}^{L}\epsilon_i ~ n_{i,\sigma}(1-n_{i,-\sigma}),\\ \nonumber
 H_U&=& U\sum_i^{L} n_{i,\uparrow}n_{i,\downarrow}.
 \end{eqnarray}
The Hamiltonian $H_K$ represents the kinetic term characterized by hopping processes between nearest-neighbor sites $\langle ij \rangle$  of the lattice. It corresponds to  Eq. (\ref{cor-hop}) with $t_{ab}=0$
and identical amplitudes for $t_a$ and $t_b$, as indicated in the sketch of Fig. \ref{hopscheme}. 
This is precisely the exactly solvable limit where
  the hopping process changing the number of occupied sites and introducing antiferromagnetic correlations is forbidden.\cite{aa1,aa2}
    In addition to the correlated hopping, there is a disorder potential characterized by local random
 energies $-W<\epsilon_i<W$ (third term) which acts on the singly occupied sites. The homogeneous case corresponds to the limit $W=0$, which is the exactly solvable case. The Hamiltonian $H_U$ describes the 
  Coulomb repulsion with $U>0$, which acts only on doubly occupied sites. 
  
 As in the case studied in  Ref.  \onlinecite{aa1,aa2} we can verify that the number of doubly occupied sites $N_d=\sum_i n_{i,\uparrow}n_{i,\downarrow}$  is conserved,  $[H, N_d]=0$. Hence, the particles can exist 
 in the lattice in the form of single fermions or doublons. The latter  are defined by  pairs of particles with different spin 
 occupying the same site. The number of each of these species is separately conserved. The total number of particles is also conserved and can be expressed as $N=N_f + 2 N_d$, where  $N_f$ is the number of the unpaired particles. The role of $U$ is equivalent to a chemical potential for the doublons.

  In the limit of 
 $W=0$ and $|t_a|=|t_b|=t$, studied in Refs. \onlinecite{aa1,aa2}, the ground state at half-filling ($N=L$) displays a Mott transition at the critical value $U_c=4t$. For $U>U_c$ the ground state corresponds to a zero energy state, where
 all the sites of the lattice are occupied by a single particle. This state is characterized by $N_d=0$,  the kinetic energy is zero ($\langle H_K \rangle=0$)
 and  is $2^L$ degenerate due to all the possible spin orientations. As it does not have any  special magnetic order  it fits to the picture of the Mott insulator. In this phase, there is a gap between the ground state and the lowest-energy excited state in the charge sector, which depends linearly in $U$. For $U<U_c$ the energy gap closes, the ground state is in a degenerate metallic phase  with states containing   superconducting order in the manyfold and it is characterized by  a
 $\langle H_K \rangle \neq 0$ and $N_d \neq 0$. 
  
 We will show in the next section that  the present model for $U=0$ can be mapped to  the usual spinless Anderson model \cite{and}. Therefore, the ground state of $H$  is clearly an Anderson gapless localized state for arbitrary small strength of the disorder $W \neq 0$. 
  Since the number of doublons does not contribute to the kinetic energy, it is also clear that for those parameters where the ground state has $N_d \neq 0$, the system may be in an Anderson-like localized phase with a finite number of doublons. This is expected to happen also when the effect of  disorder is introduced  while $U<U_c$. The question that arises is about the nature of the Mott phase as disorder is introduced for $U>U_c$. The investigation of this phase is  the main goal of the present work.

\section{Method}\label{method}
In order to find the spectrum of $H$ we follow a similar procedure to the one introduced in Refs. \onlinecite{aa1,aa2}. We focus on open boundary conditions and start by mapping the Hamiltonian 
 $H_K$ to a spinless Anderson Hamiltonian with $N_f$ particles in $L$ sites. To this end it is convenient to express the different  states of a given lattice site in terms of the following
 representation: 
 $|0 \rangle \rightarrow e_i^{\dagger} |0\rangle \equiv |\circ \rangle $, 
 $ c^{\dagger}_{\sigma} |0 \rangle \rightarrow f_{i, \sigma}^{\dagger} |0\rangle \equiv |\sigma \rangle$, 
 $ c^{\dagger}_{\uparrow} c^{\dagger}_{\downarrow} |0 \rangle \rightarrow b_i^{\dagger} |0\rangle \equiv |\bullet \rangle$. Here $f_{i \sigma}$ are fermionic  while $e_i$ and $d_i$ are bosonic operators that
 obey the following constraint
\beq
e^{\dagger}_i e_i + d^{\dagger}_i d_i + \sum_{\sigma} f^{\dagger}_{i \sigma} f_{i \sigma}=1,
\edq
which implies that a given site may have only one type of boson ($\circ$ or $\bullet$) or fermion with only one type of spin component ($\uparrow$ or $\downarrow$).
We substitute this representation in $H_K$ and focus on $|t_a|=|t_b|=t$. \cite{note} The resulting Hamiltonian reads
\beq
H_K= t \sum_{\langle i j \rangle, \sigma} \left[f^{\dagger}_{j \sigma} f_{i \sigma} \left( e^{\dagger}_i e_j + d^{\dagger}_i d_j \right) + h. c. \right] + \sum_{i \sigma} \epsilon_ i f^{\dagger}_{i \sigma}  f_{i \sigma}.
\edq
This model has two SU(2) local symmetries. The usual spin-1/2 symmetry with generators 
\beq
S_i^z= (f^{\dagger}_{i \uparrow} f_{i \uparrow} -  f^{\dagger}_{i \downarrow} f_{i \downarrow} )/2, \;\;\;\; S^+_i= f^{\dagger}_{i \uparrow} f_{i \downarrow}, \;\;\;\; S^-_i= f^{\dagger}_{i \downarrow} f_{i \uparrow},
\edq
and the $\eta$-pairing symmetry with generators
\beq
\eta_i^z = (1- \sum_{\sigma } f^{\dagger}_{i \sigma} f_{i \sigma} - 2 d^{\dagger}_i d_i)/2, \;\;\;\; \eta^+_i = e^{\dagger}_ i d_i, \;\;\;\;\; \eta^-_i = d^{\dagger}_ i e_i.
\edq
Here we notice that the states $\circ$ and $\bullet$ of the $\eta$-pairing symmetry are akin to $\uparrow$ and $\downarrow$ of the usual spin symmetry. Interestingly, we can  verify 
\beq \label{com}
\left[H_K,S_i^+\right]=\left[H_K,S_i^- \right]= \left[H_K,\eta_i^+\right]=\left[H_K,\eta_i^- \right]=0.
\edq
Therefore, we can work in the subspace corresponding  to the highest weight representation of  these SU(2) algebras. This is equivalent to working in the subspace where all the fermions have $\uparrow$ spin
and all the bosons are  $\circ$. We can diagonalize the Hamiltonian in this subspace and then, due to (\ref{com}), we know that each eigenstate $|\psi_m\rangle$ will be degenerate with states resulting from the
application of all the  lowering operators
$S_i^-|\psi_m \rangle$ and $\eta_i^- |\psi_m \rangle$. 

For a fixed number of particles $N$, and a given number of doublons $N_d$, these eigenstates have a degeneracy  
 $2^{N_f} \times C(N-N_f, N_d)$, with $C(N-N_f,N_d)$ being the combinatory number. This is  due to the different spin orientations of the unpaired particles and the different  possibilities  for allocating the doublons in the
 $N-N_f$ lattice sites which are not occupied by the unpaired fermions.
 The eigenenergies of $H$ for $U=0$ only depend on $N_f$. More precisely, these eigenenergies for $U=0$ are those of $N_f$ spinless fermions in a lattice with $L$ sites and a disorder potential profile $\epsilon_i$.  For $U \neq 0$ we notice that the Coulomb interaction acts like a chemical potential for the doublons. Then, we must add to the eigenenergies of $H_K$ the quantity $U N_d$.

The states of the basis with highest weight can be mapped to the states of $N_f$ spinless fermions. Therefore, the problem of diagonalizing the Hamiltonian $H_K$ can be mapped to diagonalizing
the  Anderson Hamiltonian in $L$ sites with $N_f$ spinless particles, which can be represented by a tridiagonal matrix with diagonal elements $\epsilon_i, \; i=1, \ldots, L$ and band elements $t$.
The single-particle eigenenergies of this problem are the eigenvalues of that matrix, $e_j, \; j=1, \ldots, L$, and the $N_f$-particle energies are
\beq
E^{(N_f)}_m= \sum_{j=1}^{N_f} e_{m(j)},
\edq
with $m=1, \ldots C(L, N_f)$. These correspond to fill in $N_f$ of the $L$ single-particle states, labelled with $m(j), \; j=1, \ldots, N_f $, with only one particle, in consistency with the Pauli principle.
The eigenenergies of the $N$- particle states with $N_f$ single fermions and $N_d$ doublons  are easily obtained for any value of $U$ from
\beq \label{en}
E^{(N)}_m= E^{(N_f)}_m + U N_d, \;\;\;\; N=N_f+ 2 N_d.
\edq
Hence, all the eigenenergies for the system with $N$ particles can be obtained by considering all the eigenenergies (\ref{en}) with all combinations of numbers of free particles and doublons ranging from
$N_f=0,\; N_d=N/2$ to $N_f=N,\; N_d=0$, satisfying the constraint of adding to a total number of particles $N$ as indicated in (\ref{en}).
\begin{figure}[h!]
\begin{center} 
\includegraphics[width=0.48\textwidth]{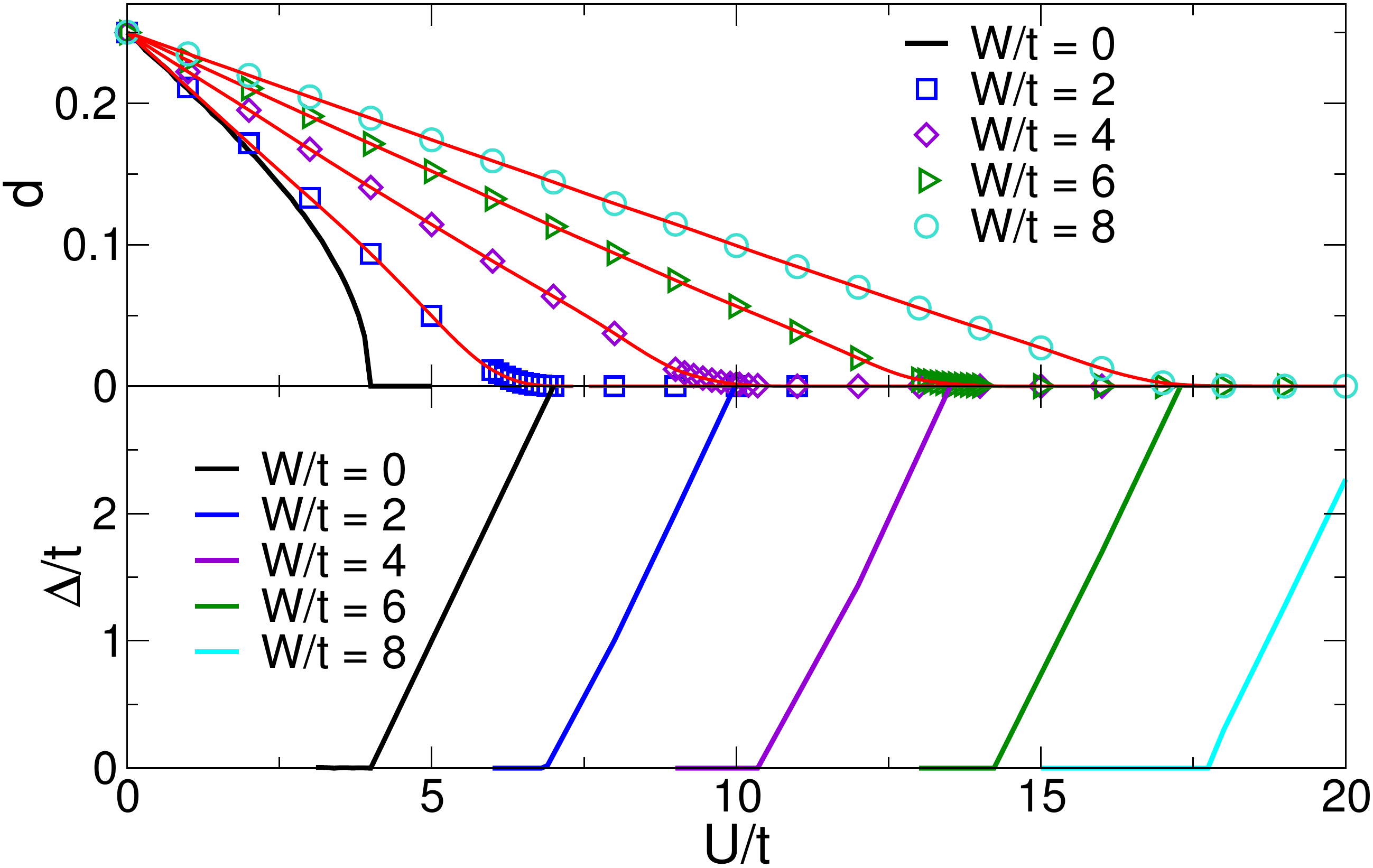}
\caption{ (Color online)  Density of doublons $d$ (top) and gap between the ground state and the first excited state, $\Delta$, (bottom) as functions of  the Coulomb interaction $U$ for different values of disorder potential $W=0,2,4,6,8$.
The two quantities are averaged over 2000 realizations of disorder in a lattice with $L=N=1200$. 
All the energies are expressed in units of the hopping parameter $t$.
}\label{fig1}
\end{center}
\end{figure}

In order to investigate the phase diagram of this model at hall-filling ($N=L$) we follow the procedure described above in chains with different size  $L$ for several values of $U$ and 2000 realizations of the
local energies $\epsilon_i$, which are randomly distributed with equal probability within the interval $(-W, W)$. We focus on the charge gap between the ground state and the first charge excitation 
\beq
\Delta=E_{1} - E_{0}, 
\edq 
where $E_{0}$  and $E_1$ are, respectively,  the ground state energy  and the first excited state  of the $N$ particles in the $L$-site chain. Typically,  the  two associated eigenstates 
 differ in one  doublon. We also analyze 
the average over  the density of doubly occupied sites 
\beq
d=N_d/L.
\edq
 To evaluate the phase diagram, we perform finite size scaling with sizes $L \leq 1200$ and extrapolate the  results to the thermodynamic limit.
 
\section{Results}\label{results}
As discussed in Section \ref{model} we do no expect any metallic phase at half-filling $N=L$ in the present model when $W\neq 0$. For $U=0$ and finite $W$ this model is equivalent to the Anderson model, which
is always in a localized phase in 1D. In the limit where $W=0$, the system is in a metallic phase for $U<4t$, characterized by a finite density of doublons $d \neq 0$ and a vanishing value of the charge gap 
$\Delta=0$.
Instead, for $U>4t$ the ground state is in the Mott-insulator phase with $d=0$ and a finite energy gap $\Delta$. 
 The ground state energy for $U=W=0$ is in the subspace with  $N_f=N/2$ and $N_d=N/4$. As $U$ increases  keeping $W=0$, $d$ decreases and vanishes at the critical value $U=4t$, where the Mott transition takes place, while, in the in the thermodynamic limit where $L \rightarrow \infty$, $\Delta =0$.
 In the Mott-insulator  phase, for $U> 4t$,  $\Delta$ becomes finite.
 
In what follows, we analyze the behavior of the density of doublons $d$ and $\Delta$ in the presence of disorder. The average over $2000$ disorder realizations of these two quantities is shown in Fig. \ref{fig1} for a finite-size lattice. 
For $W=0$ we can distinguish the characteristics of the Mott-transition above described. The (small) finite value of $\Delta$ for $U<4t$ is due to finite-size effects and it extrapolates to zero as the lattice size 
$L \rightarrow \infty$.  For $W \neq 0$ within the range $U<4t$ we see the same qualitative behavior for $d$ and $\Delta$. We identify the phase within this region of parameters with Anderson localization,
as an extension of the limiting case where $U=0$ and $W \neq 0$.  A dramatic change in the behavior of the two quantities shown in Fig. \ref{fig1} is, instead,  observed as a function of $W$ starting from the
Mott-insulator phase at $U>4t$. We see that the averaged density of doublons $d$ increases from zero while the $\Delta$ decreases as  $W$ increases. 
\begin{figure}[h!]
\begin{center}
\includegraphics[width=0.48\textwidth]{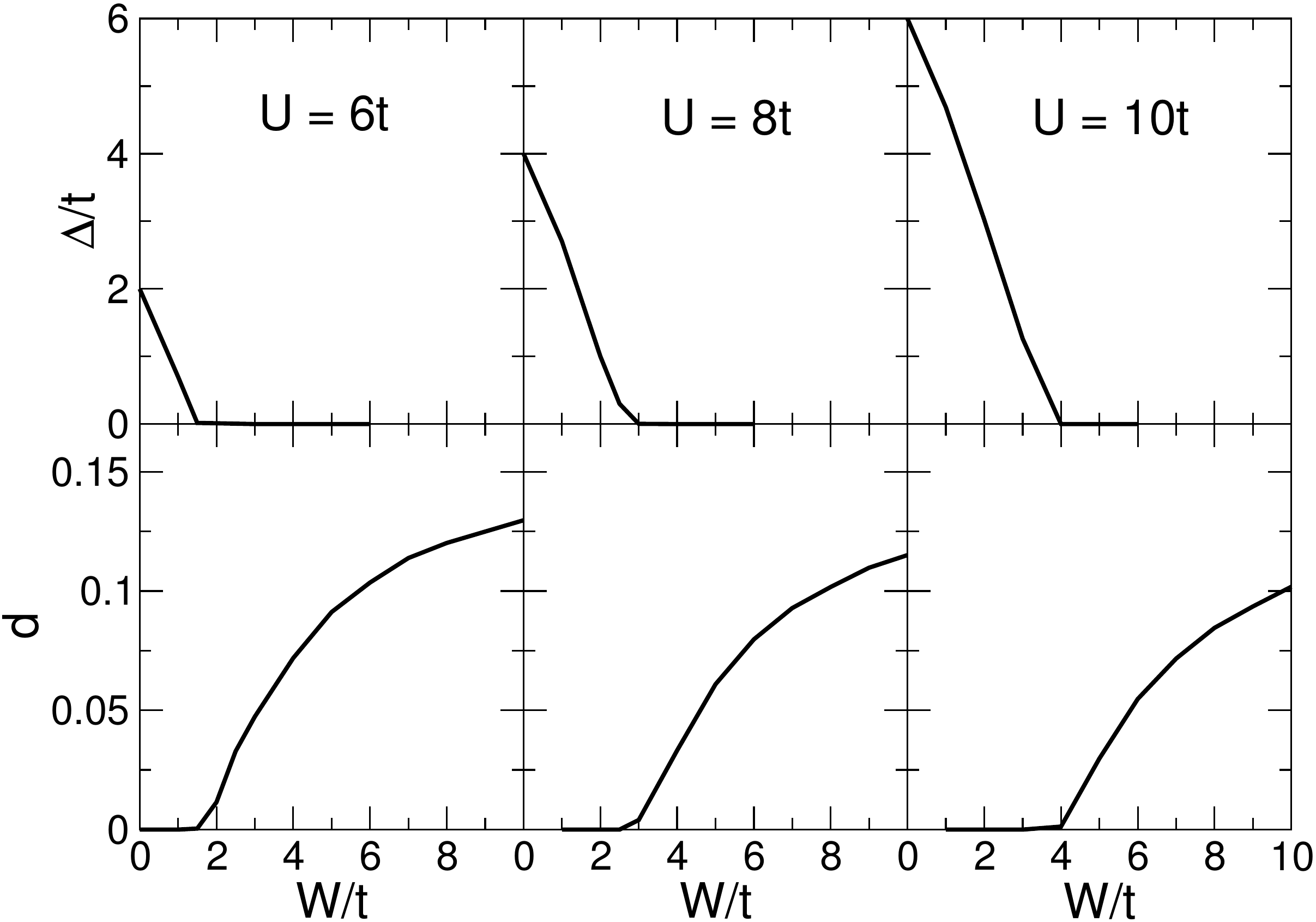}\caption{(Color online) Top panels. Energy gap value as a function of the parameter $W$ for  $U=6t$ (left), $U=8t$ (center), $U=10t$ (right). Bottom panels. Disorder-averaged 
density of doublons $d=N_d/L$ for the same values of $U$. All these data correspond to extrapolations to the thermodynamic limit from chain sizes $L \leq 1200$ and averaging over 2000 realizations of disorder.  }
\label{fig3}
\end{center}
\end{figure}
In Fig. \ref{fig3} we show  the behavior of $\Delta$ and $d$ as functions of $W$, for fixed values of the Coulomb interaction $U$ . We focus on the regime with $U>4t$. We study chains of different sizes and extrapolate
to the thermodynamic limit.   These results indicate the existence of a critical value $W_c$, such that for $W<W_c$ the density of doublons is vanishingly 
small and extrapolates to zero as $1/L \rightarrow 0 $, while for $W>W_c$ the density of doublons becomes finite and increases as a function of $W$. This change in the behavior of the density of doublons as a function
of $W$ is accompanied by a change in the behavior of $\Delta$ as a function of $W$. In fact, for $W<W_c$ we find that $\Delta$ extrapolates to a finite value in the thermodynamic limit,
while $\Delta \rightarrow 0$ for $W > W_c$. These features are consistent with a transition from a Mott-insulator phase to an Anderson-localized phase at $W_c$. 

\begin{figure}[h]
\begin{center}
\includegraphics[width=0.45\textwidth]{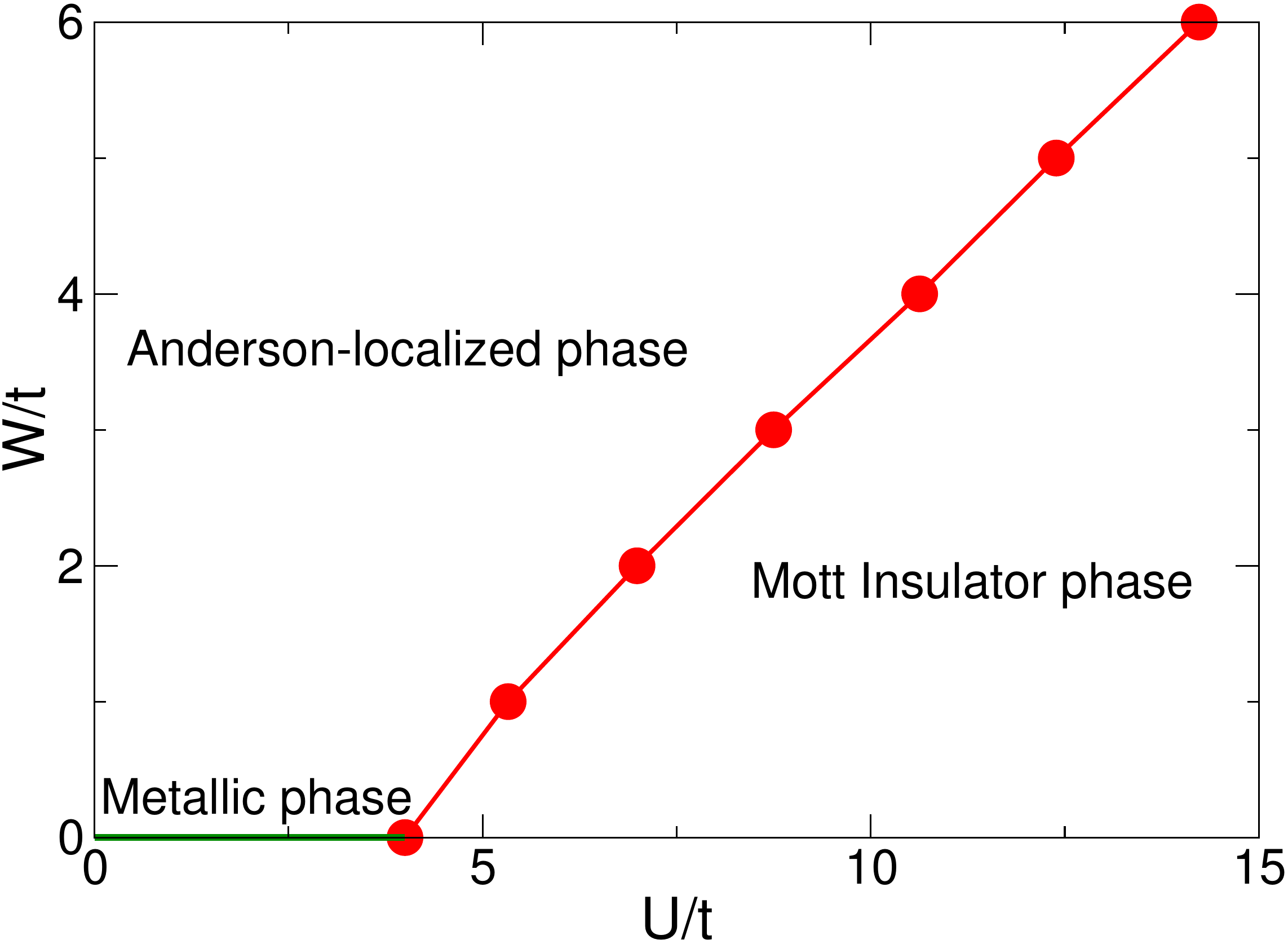}\caption{(Color online) 
Phase  diagram of the disordered Hubbard model with correlated hopping at half-filling $N=L$ in 1D.  The phases are:  (i) metallic phase for $W=0$ and $U<4t$, (ii) Anderson-localized phase for $W\neq 0$ and $U<4t$ as well as for $W>W_c$ and $U>4t$  and 
(iii) a Mott insulator phase for $W<W_c$ and $U>4t$.  The phases (i) and (ii) are characterized by $d\neq 0$, and $\Delta=0$. 
The phase (iii) is characterized by $d=0$, and $\Delta \neq 0$. The line separating the Mott insulator from the Anderson-localized phases corresponds to the
$\Delta=0$  and $d=0$  in the thermodynamic limit. 
The size of the symbols is proportional to the uncertainty on the numerical data. }\label{fig4}
\end{center}
\end{figure}

The inferred phase diagram is shown in Fig. \ref{fig4}.  The line separating the Mott insulator from the Anderson-localized phase is evaluated calculated from the criterion of vanishing $\Delta$ and vanishing $d$ in the thermodynamic limit. Polynomial extrapolations of $\Delta$ and $d$ as functions of $1/L$ have been carried out with sizes up to $L=1200$. The two estimates agree within the numerical precision. 
The same procedure can be followed to get the exact solution away from half-filling. In this case, the system is in the metallic phase for $W=0$ for any value of $U$ while it localizes for an arbitrary small strength of  $W$. 

Before closing this section, we briefly comment on similarities and differences between the phase diagram of disordered Hubbard model with correlated hopping in 1D studied in the present work
 and the phase diagram of the disordered Hubbard model at half-filling in higher dimensions described by
dynamical mean field theory  (DMFT).  In the latter case, a phase diagram with the same  phases identified in Fig. \ref{fig4} has been derived. \cite{jan-vol,ulm,sem}  In fact, the
 main underlying   characteristic shared by  the model we study here
and the DMFT description of the Hubbard model is the absence of antiferromagnetic correlations. In our case, this is an intrinsic property of the model, while in  the mean-field solutions this is encoded in the approximation of the momentum-independent self-energy. In the present 1D case, the metallic phase is confined to $W=0$, while in the higher dimensional DMFT cases, it also extends to a region with finite $W$. 
It is also interesting to stress that   the Mott phase can be identified by the change in the behavior of the doubly occupancy. The doubly occupancy as a good order parameter to characterize the transition from a metallic
to the Mott-insulating phase,
akin to the magnetization in the Ising model, was introduced in Ref. ~\onlinecite{cas}. This idea was also followed in the Landau theory for  the DMFT-Mott transition in Ref. ~\onlinecite{lange}, where
the Mott-insulating phase is characterized by a few number of doubly occupied sites, while the metallic one is identified as a liquid rich in doubly occupied sites. This is precisely the case of the model studied here. In our case,  the fact that the Mott-insulating phase is defined by an exactly vanishing number of doublons is due to the fact that the number of doubly occupied sites is a conserved quantity in the correlated-hopping Hamiltonian.

\section{Conclusions}\label{conc}
We have analyzed the phase diagram of the correlated Hubbard model with disorder at half-filling in 1D. The different phases are summarized in Fig. \ref{fig4}. Without disorder the model has a metallic and a Mott-insulator phase. Our results indicate that the metallic phase  becomes unstable and localizes as in the Anderson model for an arbitrary small disorder strength. This phase is characterized by a vanishing gap in the charge-excitations and a finite density of doublons in the ground state. Instead, the Mott-insulator phase, which is characterized by a vanishing density of doublons in the ground state and a finite charge gap, becomes stable up to a critical strength of disorder, where a phase transition to an Anderson-localized state takes place . 
The possibility of clearly identifying these two insulating phases makes this model appealing for further theoretical studies and also to be realized in optical lattices and cold atoms. Several interesting issues remain to be further investigated in the future. In particular, the possible emergence of a metallic phase  for finite disorder strength in systems of higher dimensionality,
and the role of antiferromagnetic and charge-density wave correlations, which could be  introduced  by means of an extra correlated hopping process and nearest-neighbor Coulomb interactions, generalizing the model of Refs. \onlinecite{aga1,aga2} with the addition of disorder. 

While the experimental realization of the disordered correlated hopping model in the limit we studied here is not obvious in solid state real materials, its implementation in optical lattices is within the scope of current experiments. \cite{exp1,exp2,exp3} Different  mechanisms for the experimental implementation  of disorder in these systems have been  reported. \cite{dis1,dis2,dis3} In particular, in Ref. \onlinecite{dis3}, the disorder has been introduced by means of localized impurity atoms. In the case that the latter are spin polarized, they would magnetically couple only to the singly occupied sites of the lattice under investigation, which would correspond a realization of the type of disorder we are considering in he present work.

\section{Acknowledgements}
We acknowledge support by CONICET, MinCyT and UBACYT, Argentina. LA thanks the hospitality of the ICTP-Trieste as well as the support of a Simons associateship.


\begin{thebibliography}{99}
\bibitem{note} As already noticed in Ref. \onlinecite{aa1} the sign of $t_a$ and $t_b$ is irrelevant in a bipartite lattice, since it is possible to change the
sign of $e_i$ and or $b_i$ in one of the sublattices, which implies a change of sign in these hopping processes.
\end{thebibliography}
\end{document}